# γ-ray production cross sections in proton induced reactions on $^{nat}$Mg, $^{nat}$Si and $^{nat}$Fe targets over the proton energy range E = (30 – 66) MeV

W. Yahia-Chérif[1], S. Ouichaoui[1,*], J. Kiener[2], V. Tatischeff[2], E. Lawrie[3], J.J. Lawrie[3], A. Belhout[1], H. Benhabiles[4], T.D. Bucher[3], A. Chafa[1], S. Damache[5], M. Debabi[1], I. Deloncle[2], J.L. Easton[3,6], C. Hamadache[2], F. Hammache[7], P. Jones[3], B.V. Kheswa[3,8], N. Khumalo[6], T. Lamula[6], S.T.H. Majola[3,9], D. Negi[3], J. Ndayishimye[3,8], S.P. Noncolela[3,6], D. Moussa[1], R. Nchodu[3], P. Papka[3,8], N. de Sereville[7], J.F. Sharpey-Schafer[6], O. Shirinda[3], M. Wiedeking[3] and S. Wyngaardt[8]

[1] Université des Sciences et de la Technologie H. Boumediène (USTHB), Faculté de Physique, Laboratoire SNIRM, B.P. 32 El-Alia, 16111 Bab Ezzouar, Algiers, Algeria
[2] Centre des Sciences Nucléaire et des Sciences de la Matière (CSNSM), CNRS-IN2P3 et Université de Paris-Sud, 91405 Orsay Campus, France
[3] iThemba LABS, National Research Foundation, PO Box 722, Somerset West, South Africa,
[4] Université de Boumerdès, Faculté des Sciences, Département de Physique, Boumerdés, Algeria
[5] Centre de Recherches Nucléaires (CRNA), 2 Bd Frantz Fanon, Alger-gare, Algiers, Algeria,
[6] University of the Western Cape, Private Bag X17, Bellville 7535, South Africa
[7] Institut de Physique Nucléaire (IPN), CNRS-IN2P3 et Université Paris-Sud, 91405 Orsay Campus, France
[8] University of Stellenbosch, private Bag X1, 7602 Matieland, South Africa
[9] University of Cape Town, Private Bag, 7701 Rondebosch, South Africa

**Abstract:**

γ-ray excitation functions have been measured for 30, 42, 54 and 66 MeV proton beams accelerated onto C +O (Mylar), Mg, Si, and Fe targets of astrophysical interest at the separate-sector cyclotron of iThemba LABS in Somerset West (Cape Town, South Africa). A large solid angle, high energy resolution detection system of the Eurogam type was used to record γ-ray energy spectra. Derived preliminary results of γ-ray line production cross sections for the Mg, Si and Fe target nuclei are reported and discussed. The current cross section data for known, intense γ-ray lines from these nuclei consistently extend to higher proton energies previous experimental data measured up to $E_p$ ~ 25 MeV at the Orsay and Washington tandem accelerators. Data for new γ-ray lines observed for the first time in this work are also reported.

[*] souichaoui@gmail.com

## 1. Introduction:

Gamma-ray astronomy uses γ-ray lines produced in interactions of accelerated ions through the matter of astrophysical sites as a tool for probing various non-thermal processes in the Universe [1,2]. Corresponding laboratory experimental data are then essential in connection with the modeling of the γ-ray emission from these sites and for supporting gamma astronomy studies [3,4,5]. However, the examination of the literature reveals that γ-ray line production cross section data for accelerated light charged particles interacting with target nuclei of astrophysical concern are rather scarce or even lacking for projectile velocities higher than ~ 25 MeV/amu. Such nuclear data are especially needed for simulating the violent nuclear collisions taking place, e.g., in solar flares and in the interstellar medium during the encounter of low energy cosmic ray protons and alpha particles with the nuclei of most abundant elements (C, N, O, Ne, Mg, Si and Fe). In this context, we have measured nuclear γ-ray line production cross sections for accelerated protons undergoing inelastic reactions with C +O (Mylar), Mg, Si, and Fe solid targets, these elements being abundant in the solar photosphere and corona as well as in the ISM. Preliminary results for the Mg, Si and Fe targets are reported here and discussed in comparison to few available experimental data [6,7,8,9] and to the predictions of the Talys nuclear reaction computer code [10].

## 2. Experiment:

The experiment has been carried out during 4 week-ends at the separate-sector cyclotron of iThemba LABS in Somerset West (Cape Town, South Africa) on the line of the AFRODITE reaction chamber devoted to γ-ray spectroscopy measurements [11, 12]. A large solid angle γ-detection array of 8 clovers of the Eurogam-type [13] was used. Each clover contained 4 BGO Compton-suppressed, high energy resolution HP-Ge crystals. The detectors were placed at angles of 90° and 135° around the reaction chamber at a distance of 19.6 cm from its center. Proton beams of energies $E$ = (30, 42, 54, 66) MeV and current intensities ranging from ~2 up to 5 nA were accelerated under high vacuum onto $^{12}$C, $^{16}$O (Mylar), Mg, Si, Al, Ca and Fe targets. Note that only solid targets (natural or isotopic ones) were used in this experiment. They were prepared at iThemba LABS with area density thickness values ranging between 5.0 and 8.5 mg/cm$^2$. The energy calibration of the detection system was done by using standard radioactive sources of $^{137}$Cs, $^{60}$Co and $^{152}$Eu covering the γ-ray low energy range up to 1408 keV, and the 6.129 MeV line of $^{16}$O and associated 1$^{srt}$ and 2$^{nd}$ escape lines clearly generated in the γ-spectra from the Mylar target. The γ-ray detection efficiencies were measured by means of the preceding radioactive sources. For the higher energy region of the γ-spectra, efficiency curves were generated via Geant-4 simulation, then they were normalized to the measured low energy data. Finally, γ-ray backgrounds without beam on targets were systematically measured and subtracted from the recorded γ-ray spectra of interest.

### 3. Results and discussion:

A typical **γ-**ray energy spectrum measured for 54 MeV protons incident on the Si target is reported in figure 1. The numbers of counts in the γ-ray lines of interest have been extracted by a careful analysis of the recorded γ-ray energy spectra using the CERN ROOT program [14]. This allowed us to determine the production cross sections for various lines generated in the $^{nat}$Mg(p,xγ), $^{nat}$Si(p,xγ) and $^{nat}$Fe(p,xγ) proton inelastic scattering processes. Derived preliminary cross section results for known, intense (henceforth called *main*) γ-ray lines of the $^{24}$Mg, $^{28}$Si and $^{56}$Fe nuclei (the 1369 keV, 1779 keVand 847 keV lines, respectively) are reported in figures (2-4) where the cross section experimental data are compared to previous ones [6,7,8,9] and to the predictions of the Talys nuclear reaction code [10]. Two types of cross section values have been calculated and are depicted in figures (2-4) by considering: (i) only the main γ-ray line generated in proton inelastic scattering off the most abundant isotope in each used natural target (e.g. $^{24}$Mg) and (ii) that main line and lines from the (p,pnγ) and/or (p,p2nγ) reactions induced by incident protons on the minor isotopes of the same target (e.g. $^{25}$Mg and $^{26}$Mg). One can notice, first, that our data appear to extend in a consistent way to higher proton energies the cross section data for the main lines measured previously at the Orsay [6,7] and Washington [8] tandem accelerators, while the LNL cyclotron data from reference [9] for the Mg and Fe targets clearly lie above both our data and Talys code-generated values. As can also be seen in figures (2,3), one must take into account close-energy lines from minor isotopes in the $^{nat}$Mg and $^{nat}$Si targets (respectively $^{25, 26}$Mg and $^{29, 30}$Si) together with the 1369 keVand 1779 keV main lines of $^{24}$Mg and $^{28}$Si to improve the agreement between calculated and measured respective cross sections. In contrast, a fairly good agreement of experimental data with Talys calculation is observed in figure 4 for the 847 keV main line of $^{56}$Fe taken alone. Besides, new γ-ray lines for several target nuclei were observed for the first time in this experiment. Corresponding γ-ray production cross section experimental data in the case of the $^{24}$Mg and $^{56}$Fe nuclei are reported in figures (5, 6), where they are compared only to values calculated by the Talys code, in the absence of any previous counterpart experimental data. It is then worthwhile noting that the reported experimental data turn out to be very useful for constraining nuclear reaction models. Thus, e.g., the agreement of Talys-calculated cross sections with experimental data could be further improved by adjusting nuclear properties, like the couplings chemes of collective levels [6] and the parameters of optical model potentials [7], in the γ-emitting nuclei. On the other hand, the measured γ-ray line spectra and production cross sections are of great importance for comparisons to observational data from satellite missions and for the modeling of nuclear *γ*-ray line emission in various astrophysical sites [3,4,5] like, e.g., the Sun for solar flares and the interstellar medium permanently bombarded by cosmic rays.

**Acknowledgements** :

This work was carried out in the framework of a joint Algerian, South African and French scientific cooperation project. The latter is funded within a 2014-Agreement between the DGRSDT of Algeria and the NRF of South Africa while financial support was granted to the French researchers by the CSNSM and IPN of Orsay (CNRS-IN2P3 and University of Paris-Sud). Financial support has also been granted to Algerian researchers in 2012 by the USTHB University prior to the DGRSDT-NRF official Agreement.

Deep thanks are due to all concerned persons from these institutions.

**Figure captions:**

Fig. 1: typical **γ-**ray spectrum measured for 54 MeV protons scattered off the $^{nat}$Si target. The 511 keV line from electron-positron annihilation is outgrowing in the lower energy part.

Fig. 2: Experimental and Talys-calculated $\gamma$-ray production cross sections for the $^{nat}$Mg (p,x$\gamma_{1369\ kev}$) reaction (see text).

Fig. 3: Same as Fig. 2 for the $^{nat}$Si(p, x$\gamma_{1779\ kev}$) reaction

Fig. 4: Same as Fig. 2 for the $^{nat}$Fe(p, x$\gamma_{847\ kev}$) reaction

Fig. 5: Experimental and Talys-calculated $\gamma$-ray production cross sections for the $^{nat}$Mg(p, x$\gamma_{2764kev}$) reaction (new 2764 keVline see text)

Fig. 6: Same as Fig. 5 for the $^{nat}$Fe(p, x$\gamma_{477\ kev}$) reaction (new 477 keV line)

**Figures:**

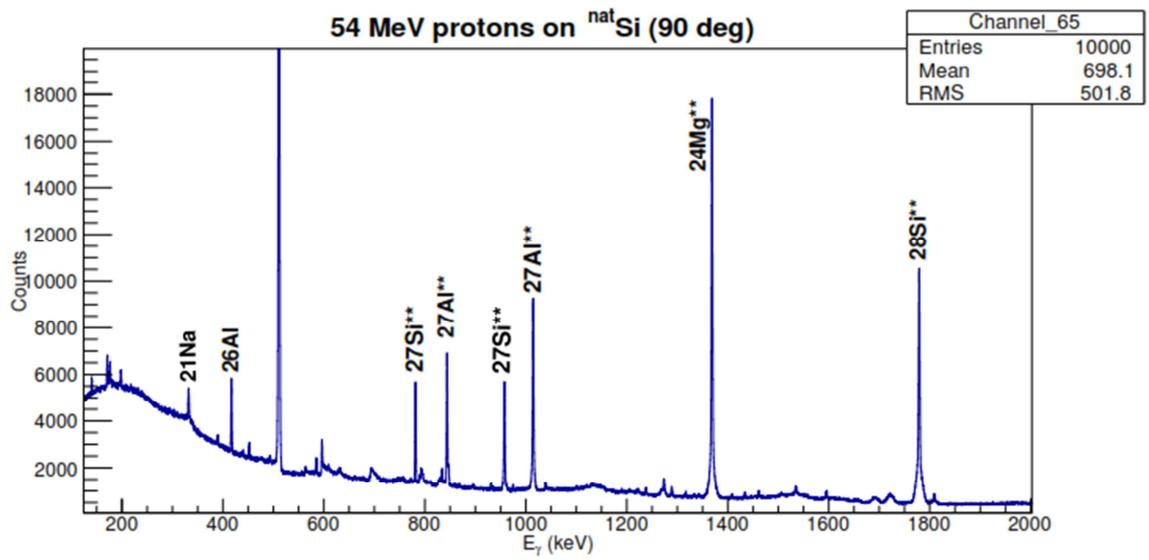

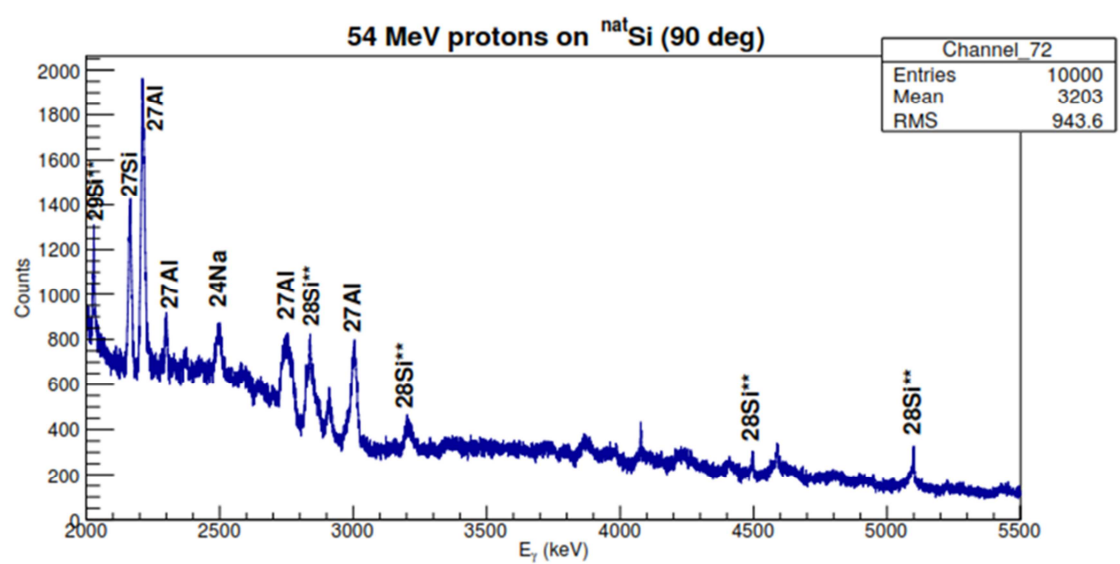

Fig.1

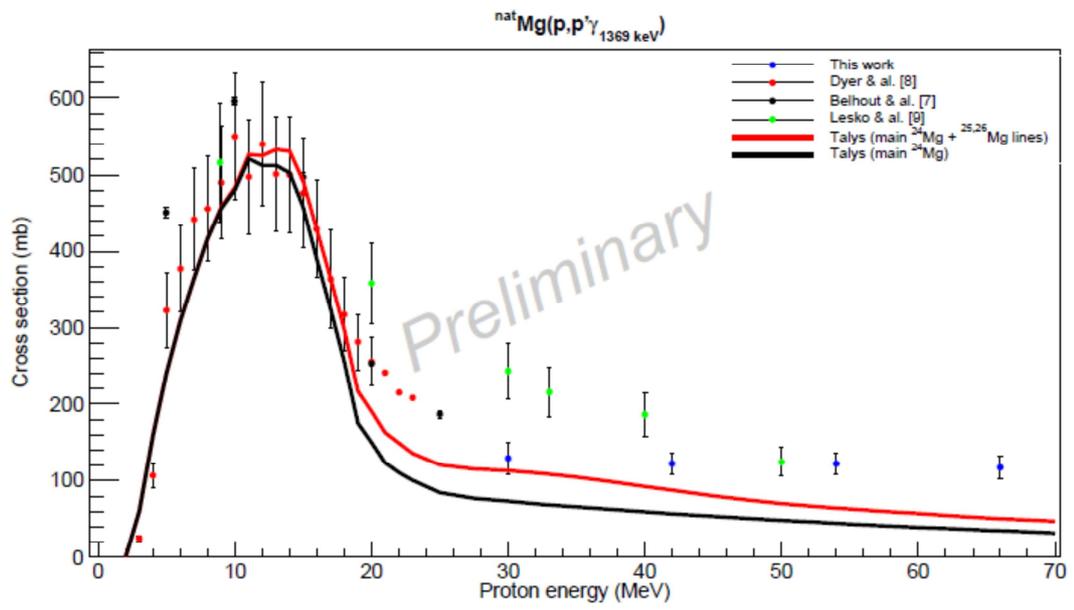

Fig.2

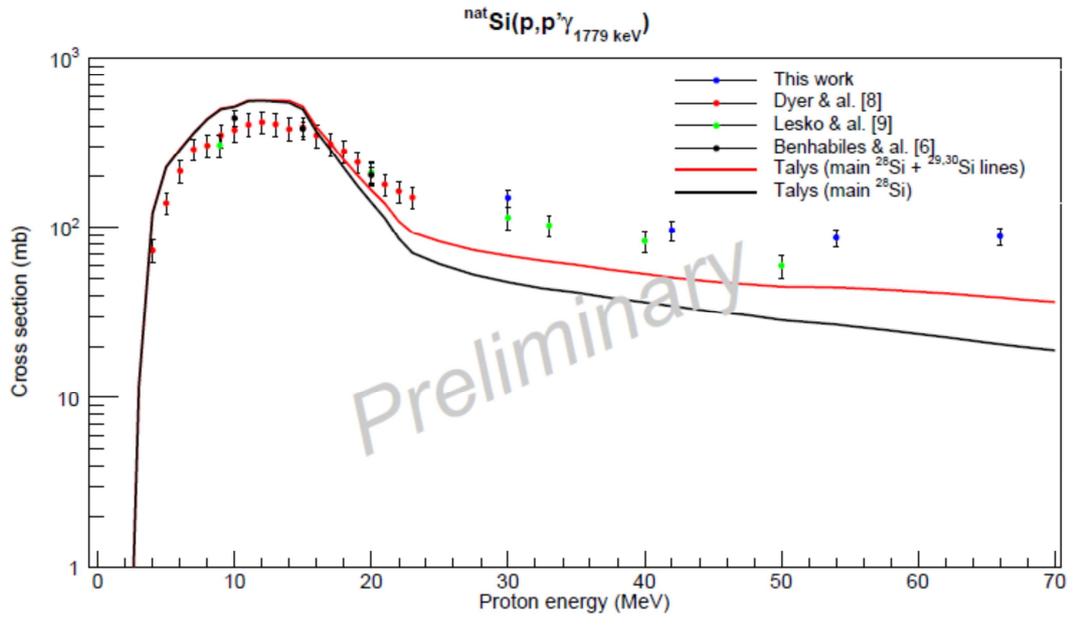

Fig.3

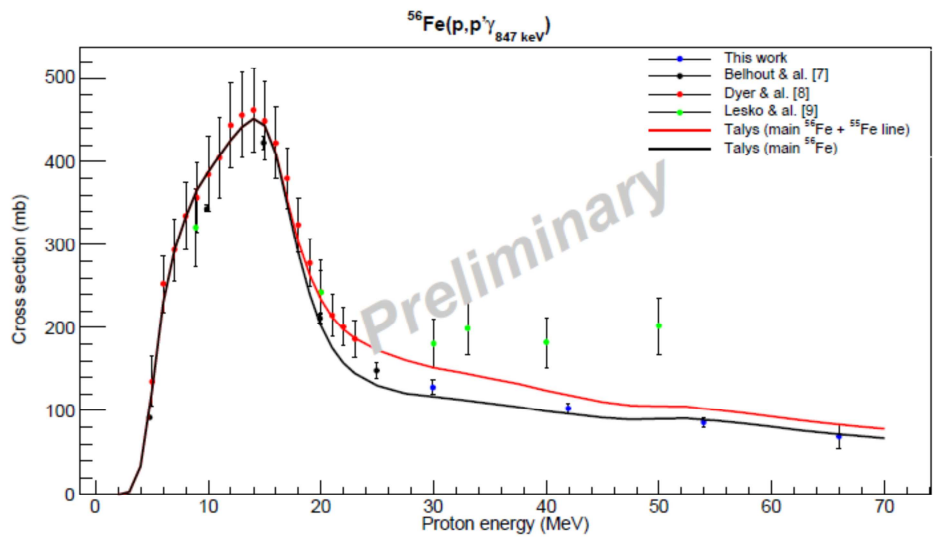

Fig.4

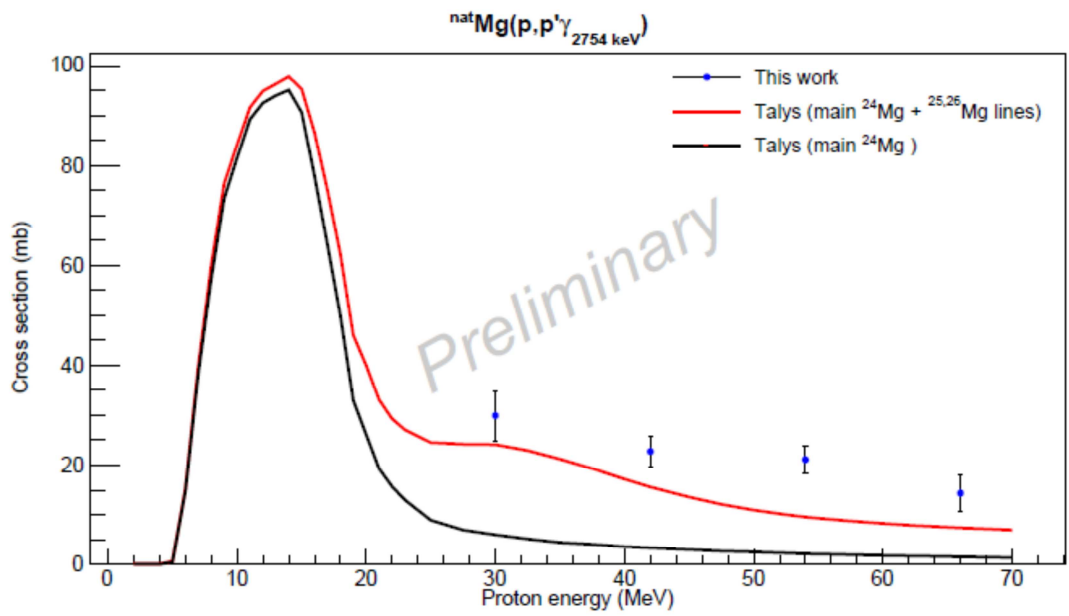

Fig.5

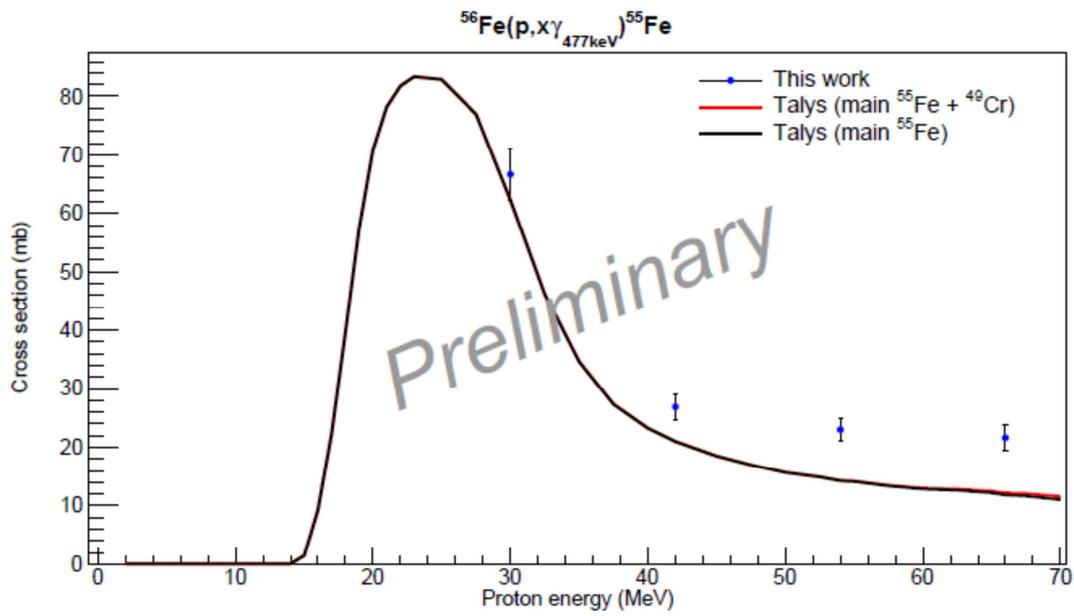

Fig.6